# Earthquake Distance and Magnitude Estimation via Calibrated Microwave Frequency Fiber Interferometry

Stavros Deligiannidis[(1)], Yuhan Wang[(2)], Christos Simos[(3)], Iraklis Simos[(4)], Andreas Fichtner[(2)], Nikolaos S. Melis[(5)], Charis Mesaritakis[(6)], Adonis Bogris[(1)]

(1) Department of Informatics and Computer Engineering, University of West Attica, Aghiou Spiridonos, 12243, Egaleo, Athens, Greece, sdeligiannid@uniwa.gr, abogris@uniwa.gr
(2) Department of Earth and Planetary Sciences, ETH Zurich, Zurich, Switzerland, yuhawang@ethz.ch, andreas.fichtner@eaps.ethz.ch
(3) Department of Physics, University of Thessaly, 35100, Lamia, Greece, christos.simos@uth.gr
(4) Department of Electrical and Electronics Engineering, University of West Attica, Aghiou Spiridonos, 12243, Egaleo, Athens, Greece, simos@uniwa.gr
(5) National Observatory of Athens, Institute of Geodynamics, Athens, Greece, nmelis@noa.gr
(6) Department of Biomedical Engineering, University of West Attica, Aghiou Spiridonos, 12243, Egaleo, Athens, Greece, cmesar@uniwa.gr

**Abstract** *We present a calibration framework for a Microwave Frequency Fiber Interferometer (MFFI) aimed at estimating earthquake distance and magnitude, with the long-term goal of enabling early warning capabilities. This marks the first demonstration of calibrated MFFI sensing for quantitative seismic parameter retrieval. ©2025 The Author(s)*

**Introduction**

Fiber-optic infrastructure is increasingly recognized as a transformative medium for large-scale environmental sensing, particularly in seismically active regions. Distributed Acoustic Sensing (DAS) has emerged as a leading technology, converting optical fibers into dense arrays of virtual seismometers, capable of high-resolution strain-rate measurements over distances up to around 100 km [1]. Recent studies have explored DAS amplitude fidelity and its capabilities for Earthquake Early Warning (EEW) systems, revealing its potential for ground motion estimation and subsurface imaging despite known challenges such as saturation and cable-ground coupling uncertainty [2-3]. However, it does not support the interrogation of long-haul transoceanic cables, thereby limiting its ability to offer early warning for earthquakes at larger distances from the coast. Complementary methods, such as polarization sensing over live optical networks [4], offer additional capabilities in network health monitoring and earthquake detection, but have been largely unexplored as calibrated systems for the estimation of earthquake magnitude and distance.

MFFI represents a promising alternative to DAS and polarization-based techniques, offering robust, high-sensitivity phase-based strain measurements compatible with long-haul telecom fibers. Unlike optical interferometry, which demands ultra-stable laser sources [5], MFFI leverages commercially available sub-Hz microwave oscillators, reducing cost and complexity without compromising sensitivity [6]. In prior work, we demonstrated the feasibility of MFFI in submarine environments, including the detection of earthquakes with magnitudes as low as 1.5, ocean tides, and sea-state dynamics over a 15.6 km fiber link between Cephalonia and

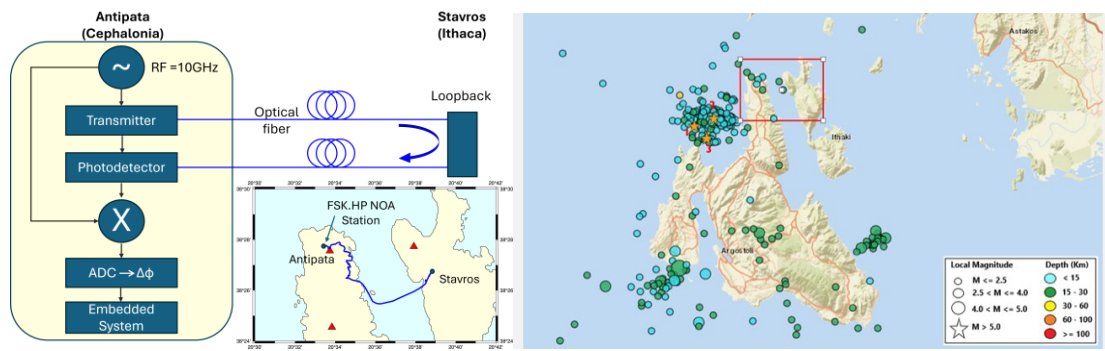

**Fig. 1:** a) MFFI building blocks and the map showing the cable route in land and sea areas. The University of Patras seismic at Fiskardo (FSK.HP) is also depicted in the map (red triangle), south close to the fiber in blue. b) Local seismicity during the experiment located by NOA (red box the experiment area).

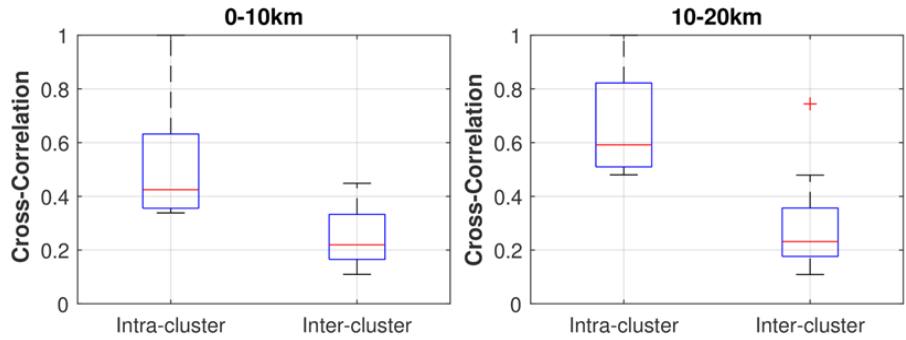

**Fig. 2:** Boxplots of intra-cluster and inter-cluster cross-correlation coefficients for the 0–10 km and 10–20 km distance clusters. Intra-cluster correlations (within each group) are significantly higher than inter-cluster values, demonstrating clear separation and strong internal consistency in the near-fault event clusters.

Ithaca, Greece [7]. Previous deployments have primarily focused on qualitative signal analysis or detection capability. A major limitation for real-time applications—especially for EEW systems—is the absence of a reliable framework that can derive meaningful earthquake parameters using raw interferometric data. Calibration becomes especially critical when applied to pre-existing, heterogeneous telecom infrastructure, where fiber geometry, burial depth, and coupling conditions vary significantly. This challenge is well known in the DAS community, which has developed empirical calibration schemes for amplitude and saturation thresholds to enable magnitude scaling [2],[7].

In this work, we introduce a novel, data-driven calibration framework for MFFI, enabling the first quantitative estimation of earthquake epicentral distance and magnitude based on this interrogation technique. By integrating data from conventional seismic networks and DAS systems, we derive empirical models that relate MFFI strain-acceleration data to earthquake source parameters. Our results reveal strong agreement with co-located DAS and earthquake recorded data. This work lays the foundation for implementing MFFI in real-time EEW architectures, in regions with sparse seismic instrumentation.

**Field trial and experimental setup**
The field trial was conducted on the island of Cephalonia, Greece, near the town of Fiscardo (Fig. 1a). The fiber-optic link used in the experiment was provided by the Hellenic Telecommunications Organisation S.A. (OTE) and spans a total length of 15.6 km. It consists of three segments: 7.2 km of terrestrial fiber in Cephalonia, 7.1 km of submarine cable crossing the Ionian Sea between Cephalonia and Ithaca, and 1.3 km of terrestrial fiber on Ithaca island. The MFFI system (Fig. 1a) was installed in

**Tab. 1:** Distribution of Earthquake Events across distance clusters from the fiber

| Distance from fiber (km) | Num. of Events |
|---|---|
| 0-10 | 29 |
| 10-20 | 20 |
| 20-40 | 10 |
| >40 | 16 |

Antipata and operated for two months in parallel with a Silixa iDAS v2 system, starting on 24 April 2024 [7]. The fiber link lies adjacent to the Cephalonia Transform Fault, one of the most seismically active zones of Europe (Fig. 1b). During the observation period, multiple series of earthquakes with local magnitudes ranging from ML=1.5 to 3.0 were recorded in the region. This setting provided a unique opportunity to assess the performance of MFFI against reference data from NOA seismometers, particularly the nearby seismic station FSK.HP operated by University of Patras (Fig. 1a).

**Results**
To establish a robust calibration framework for the MFFI system and assess its ability to distinguish between seismic events at varying distances, we compiled a dataset of recorded earthquakes during the two-month monitoring period. Events were cross-referenced with the NOA earthquake catalogue. To ensure both relevance and signal fidelity, only earthquakes with magnitudes > 1.6 were considered, as these exhibit higher signal-to-noise ratios (SNR). A total of 75 earthquakes met the selection criteria. Events were grouped into four clusters based on epicentral distance from the fiber, calculated using NOA catalogue coordinates and the geometric center of the fiber. P- and S-wave arrivals were not consistently visible in MFFI data, so they were not used for distance estimation. The clusters are based on the epicentral distances ranges 0–10 km, 10–20 km, 20–40 km and >40 km. The distribution of events across these clusters is shown in Table I. To extract seismic features from the raw MFFI signals, we performed a standardized preprocessing pipeline on the continuous hourly recordings. Both strain rate and strain acceleration were evaluated, with the latter exhibiting higher temporal smoothness and cluster consistency. A zero-phase Butterworth bandpass in the 1 - 7 Hz range was applied to the strain acceleration signal. This frequency band encompasses the dominant energy of local earthquakes in the Ionian region. From each detected event, a one-minute signal segment centered on the earthquake arrival was extracted and treated as a reference pattern. For each distance cluster, a representative event was selected based on its SNR and magnitude. We then computed the normalized cross-correlation coefficients between this reference and all other events within the same cluster (intra-cluster), as well as with events in different clusters (inter-cluster). To provide statistically meaningful results, the distance estimation analysis through matched filtering was focused on the first two distance clusters (0–10 km and 10–20 km) where at least 20 events are available per cluster. The results for the 0–10 km and 10–20 km clusters are illustrated in Fig. 2. Boxplots of the cross-correlation coefficients demonstrate that intra-cluster similarity is significantly higher than inter-cluster similarity, indicating strong temporal and frequency coherence among events of similar distance. Hence, with simple data driven matched filtering, a rough estimation of earthquake epicentral distance is possible.

**Tab. 2:** Normalized Mean Squared Error (NMSE) and Error Spread (Std) for Magnitude prediction across distance clusters

| Distance from fiber (km) | NMSE | Std |
|---|---|---|
| 0-10 | 0.48 | 0.25 |
| 10-20 | 0.23 | 0.15 |
| 20-40 | 0.26 | 0.34 |
| >40 | 0.47 | 0.42 |

Based on the fact that epicentral distance can be estimated by clustering and matched filtering, we proceed to assess the system's ability to estimate earthquake magnitude. In this analysis, we considered the set of 75 recorded events, extending beyond the 0–10 km and 10–20 km clusters to include all available earthquakes. To derive a robust estimator of magnitude from MFFI data, we first converted the reported magnitudes into an equivalent seismic energy representation.

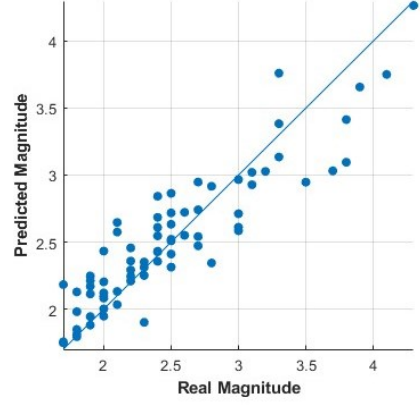

**Fig. 3:** Predicted versus real earthquake magnitudes based on the log-linear model using distance and signal variability

Each event's strength was calculated based on the amplitude of the waveform of strain acceleration minus the amplitude of background noise within the 1 - 7 Hz band. Inspired by standard magnitude definitions for seismometer data, we assumed that the magnitude $M$ is a linear function of both the logarithm of the epicentral distance $D$ and the logarithm of the event amplitude,

$$M = a \cdot \log_{10}(D) + b \cdot \log_{10}(S) + c, \quad (1)$$

with parameters $a$, $b$ and $c$. Through linear fitting of half of the measured events, we found $a$=1.364, $b$=0.7592 and $c$=6.0735. Table II reports the normalized mean square error (NMSE) and standard deviation of the prediction error for each distance. Fig. 3 illustrates the relation of magnitudes estimated by NOA seismometer and MFFI data for all events, using the fitted log-linear model. The MFFI-estimates align closely with NOA magnitudes, demonstrating good magnitude estimation performance across a magnitude range from ML=1.6 to ML=4.

**Conclusions**

We presented a calibration framework for MFFI that enables the estimation of earthquake distance and magnitude based on strain acceleration signals. A dataset of 75 seismic events was analyzed, revealing strong intra-cluster similarity and clear separability based on epicentral distance. A log-linear model incorporating signal amplitude and distance achieved magnitude estimations across a magnitude range from 1.6–4.2, with errors mostly below 0.25 magnitude units. Our results demonstrate the potential of MFFI as a low-cost, high-sensitivity solution for quantitative seismic monitoring.

## Acknowledgements

We would like to thank I. Chochliouros, C. Avdoulos, A. Apostolatos and D. Alissandratos from OTE for providing access to the fiber and assisting in MFFI and DAS installation at Antipata. The University of Patras Seismology Lab that operates the station FSK.HP. This work has been supported by Horizon Europe project ECSTATIC under grant agreement 101189595 and by ETH Zurich.